\documentclass[a4paper,11pt]{article}%
\usepackage{amsmath}
\usepackage{amssymb}
\usepackage{amsfonts}
\begin{document}

\title{\textsf{Solitons of Sigma Model on Noncommutative Space as Solitons of
Electron System}}
\author{\textsf{Hideharu Otsu } \thanks{otsu@vega.aichi-u.ac.jp}\\Faculty of Economics, Aichi University,\\Toyohashi, Aichi 441-8522, Japan
\and \textsf{Toshiro Sato} \thanks{tsato@mie-chukyo-u.ac.jp}\\Faculty of Law and Economics, Mie Chukyo University, \\Matsusaka, Mie 515-8511, Japan
\and \textsf{Hitoshi Ikemori } \thanks{ikemori@biwako.shiga-u.ac.jp}\\Faculty of Economics, Shiga University, \\Hikone, Shiga 522-8522, Japan
\and \textsf{Shinsaku Kitakado }\thanks{kitakado@ccmfs.meijo-u.ac.jp}\\Department of Physics, Faculty of Science and Technology,\\Meijo University,\\Tempaku, Nagoya 486-8502, Japan }
\date{}
\maketitle

\begin{abstract}
We study the relationship of soliton solutions for electron system with those
of the sigma model on the noncommutative space, working directly in the
operator formalism. We find that some soliton solutions of the sigma model are
also the solitons of the electron system and are classified by the same
topological numbers.

\end{abstract}

\newpage

\section{Introduction}

As is well known, the systems in the strong magnetic field can be described by
the noncommutative field theories. This picture has an application to the
solid state physics and effective theories derived from the string model also
belong to this category. Various analyses of field theories on the
noncommutative spaces are performed \cite{Harvey:2001yn}\cite{Nekrasov:1998ss}%
\cite{Gopakumar:2000zd}. In particular, nonlinear sigma model on the
noncommutative plane are investigated in detail and the structures of soliton
solutions are becoming clear \cite{Lee:2000ey}\cite{Lechtenfeld:2001uq}%
\cite{Lechtenfeld:2001aw}\cite{Lechtenfeld:2001gf}\cite{Furuta:2002ty}%
\cite{Furuta:2002nv}\cite{Otsu:2003fq}\cite{Otsu:2004fz}. The soliton
solutions that do not have the couterparts in the commutative space are also
known to exist in this model \cite{Otsu:2003fq}\cite{Otsu:2004fz}.

In this paper, we shall investigate the physical phenomena related to the
soliton solutions of the nonlinear sigma model. We consider two dimensional
spinning electron system in the magnetic field perpendicular to the plane.
Coulomb repulsion is considered to be present among the electrons. If we
restrict our consideration to the lowest Landau level (LLL), this system is
reduced to that on the noncommutative space \cite{Pasquier:2000gz}%
\cite{Lee:2001fk}\cite{Magro:2003bs}. In ref.\cite{Pasquier:2000gz}, the
Hamiltonian for the electron system has been constructed, where $O(3)\sigma$
model on the commutative space has been derived through the expansion of the
$\ast$product and the application of the soliton solution was also considered.
We shall study the soliton solution directly in the operator formalism. We
find that some soliton solutions of the sigma model on the noncommutative
space are also the solitons of the electron system. The configuration space of
the electron system can be identified with that of $O(3)\sigma$ model when
expressed in terms of the projector $P,$ thus the same topological numbers can
be used in classifying the configurations.

In section 2, we summarize the properties of the soliton solutions of
nonlinear sigma model on the noncommutative plane. In section 3, we study the
noncommutative BPS equations and their soliton solutions in the operator form
considering the Hamiltonian for the electron system. In section 4, we discuss
the classification of solitons by the new topological number. Finally, in
section 5, we conclude with summary and discussions.

\section{Nonlinear Sigma Model on Noncommutative Plane and Soliton Solutions}

In this section we summarize the main properties of soliton solutions on the
noncommutative two dimensional space. First we fix the notations. The space
coordinates obey the commutation relation%
\begin{equation}
\left[  x,y\right]  =i\theta
\end{equation}
or%
\begin{equation}
\left[  z,\bar{z}\right]  =\theta>0,
\end{equation}
when written in the complex variables, $z=\frac{1}{\sqrt{2}}(x+iy)$ and
$\bar{z}=\frac{1}{\sqrt{2}}(x-iy)$. The Hilbert space can be described in
terms of the energy eigenstates $\left\vert n\right\rangle $ of the harmonic
oscillator whose creation and annihilation operators are $\bar{z}$ and $z$
respectively,%
\begin{align}
z\left\vert n\right\rangle  &  =\sqrt{\theta n}\left\vert n-1\right\rangle
,\nonumber\\
\bar{z}\left\vert n\right\rangle  &  =\sqrt{\theta(n+1)}\left\vert
n+1\right\rangle .
\end{align}
Space integrals on the commutative space are replaced by the trace on the
Hilbert space
\begin{equation}
\int d^{2}x\Rightarrow\mathrm{Tr}_{\mathcal{H}}\mathrm{,}%
\end{equation}
where $\mathrm{Tr}_{\mathcal{H}}$ denotes the trace over the Hilbert space as
\begin{equation}
\text{$\mathrm{Tr}_{\mathcal{H}}$}\mathcal{O}=2\pi\theta\sum_{n=0}^{\infty
}\left\langle n\right\vert \mathcal{O}\left\vert n\right\rangle .
\end{equation}
The derivatives with respect to $z$ and $\bar{z}$ are defined by $\partial
_{z}=-\theta^{-1}\left[  \bar{z},\right]  $\ and $\partial_{\bar{z}}%
=\theta^{-1}\left[  z,\right]  $.

Next we turn to the nonlinear sigma model on the noncommutative space. As a
field variable we take the $2\times2$ matrix projector $P$ ($P^{2}=P$).
Lagrangian and topological charge are respectively
\begin{equation}
L=\frac{1}{2}\mathrm{Tr}_{\mathcal{H}}\left[  \mathrm{tr}(\partial_{t}%
P)^{2}\right]  -\frac{1}{\theta^{2}}\mathrm{Tr}_{\mathcal{H}}\left[
\mathrm{tr}\left(  [z,P][P,\bar{z}]\right)  \right]  \label{L-nls}%
\end{equation}
and%
\begin{equation}
Q=\frac{1}{4\pi\theta^{2}}\mathrm{Tr}_{\mathcal{H}}\left[  \mathrm{tr}\left\{
\left(  2P-1\right)  \left(  [\bar{z},P][z,P]-[z,P][\bar{z},P]\right)
\right\}  \right]  . \label{Top-1}%
\end{equation}
Energy for the static configuration is expressed as%
\begin{equation}
E=\frac{1}{\theta^{2}}\mathrm{Tr}_{\mathcal{H}}\left[  \mathrm{tr}(\left[
z,P\right]  \left[  P,\bar{z}\right]  )\right]  ,
\end{equation}
which leads to the energy bound \cite{Lechtenfeld:2001aw}%
\cite{Gopakumar:2001yw}%

\begin{equation}
E\geq2\pi\left\vert Q\right\vert .
\end{equation}
For $Q>0$, the BPS soliton equation \cite{Lechtenfeld:2001aw}%
\cite{Gopakumar:2001yw}\cite{Hadasz:2001cn} is%
\begin{equation}
(1-P)zP=0. \label{BPS-P}%
\end{equation}
Similarly, for $Q<0$ the BPS anti-soliton equation is%
\begin{equation}
(1-P)\bar{z}P=0. \label{aBPS-P}%
\end{equation}
The static configurations are classified by the topological charge
$Q=0,\pm1,\pm2,\cdots$\ and $\left\langle \mathrm{tr}P\right\rangle _{\infty
}=0,1,2$, the values of $\mathrm{tr}P$ at the boundary of the Hilbert space
\cite{Otsu:2004fz}, where $\left\langle \mathrm{tr}P\right\rangle _{\infty}$
is defined as
\begin{equation}
\left\langle \mathrm{tr}P\right\rangle _{\infty}\equiv\lim_{n\rightarrow
\infty}\left\langle n\right\vert \mathrm{tr}P\left\vert n\right\rangle .
\end{equation}

The following soliton solutions that satisfy the BPS equations are known. The
solitons that extrapolate into those on the commutative plane with
$Q=n,\ E=2\pi n$ and $\left\langle \mathrm{tr}P\right\rangle _{\infty}=1$ are
\cite{Lee:2000ey}
\begin{equation}
P=\left(
\begin{array}
[c]{cc}%
z^{n}\dfrac{1}{\bar{z}^{n}z^{n}+1}\bar{z}^{n} & z^{n}\dfrac{1}{\bar{z}%
^{n}z^{n}+1}\\
\dfrac{1}{\bar{z}^{n}z^{n}+1}\bar{z}^{n} & \dfrac{1}{\bar{z}^{n}z^{n}+1}%
\end{array}
\right)  . \label{sLee-P}%
\end{equation}
The anti-solitons extrapolating into those on the commutative plane with
$Q=-n,\ E=2\pi n$ and $\left\langle \mathrm{tr}P\right\rangle _{\infty}=1$ are%
\begin{equation}
P=\left(
\begin{array}
[c]{cc}%
\bar{z}^{n}\dfrac{1}{z^{n}\bar{z}^{n}+1}z^{n} & \bar{z}^{n}\dfrac{1}{z^{n}%
\bar{z}^{n}+1}\\
\dfrac{1}{z^{n}\bar{z}^{n}+1}z^{n} & \dfrac{1}{z^{n}\bar{z}^{n}+1}%
\end{array}
\right)  . \label{asLee-P}%
\end{equation}
The commutative limit of which are obtained by simply substituting the
c-number $z$ and $\bar{z}$ into (\ref{sLee-P}) and (\ref{asLee-P}), then they
reduce to the solitons and anti-solitons of the nonlinear sigma model
respectively. Furthermore, there exist solitons that do not have the
counterparts on the commutative space. They have $Q=n,\ E=2\pi n$ and
$\left\langle \mathrm{tr}P\right\rangle _{\infty}=1$ for soliton solutions
expressed as \cite{Otsu:2003fq}\cite{Otsu:2004fz}%
\begin{equation}
P=\left(
\begin{array}
[c]{cc}%
1 & 0\\
0 & \sum_{m=0}^{n-1}\left\vert m\right\rangle \left\langle m\right\vert
\end{array}
\right)  , \label{sOSIK-P}%
\end{equation}
$Q=-n,\ E=2\pi n$ and $\left\langle \mathrm{tr}P\right\rangle _{\infty}=1$ for
anti-soliton solutions;%

\begin{equation}
P=\left(
\begin{array}
[c]{cc}%
1-\sum_{m=0}^{n-1}\left\vert m\right\rangle \left\langle m\right\vert  & 0\\
0 & 0
\end{array}
\right)  , \label{asOSIK-P}%
\end{equation}
$Q=k+n,\ E=2\pi(k+n)$ and $\left\langle \mathrm{tr}P\right\rangle _{\infty}=0$
for soliton solutions;%

\begin{equation}
P=\left(
\begin{array}
[c]{cc}%
\sum_{m=0}^{k-1}\left\vert m\right\rangle \left\langle m\right\vert  & 0\\
0 & \sum_{m=0}^{n-1}\left\vert m\right\rangle \left\langle m\right\vert
\end{array}
\right)  \label{newS}%
\end{equation}
and $Q=-(k+n),\ E=2\pi(k+n)$ and $\left\langle \mathrm{tr}P\right\rangle
_{\infty}=2$ for anti-soliton solutions;
\begin{equation}
P=\left(
\begin{array}
[c]{cc}%
1-\sum_{m=0}^{n-1}\left\vert m\right\rangle \left\langle m\right\vert  & 0\\
0 & 1-\sum_{m=0}^{k-1}\left\vert m\right\rangle \left\langle m\right\vert
\end{array}
\right)  . \label{newAS}%
\end{equation}

For the finite energy configuration, the topological charge $Q$ can be
rewritten after some calculations into the following simple form
\begin{equation}
Q={\frac{1}{2\pi\theta}}\mathrm{Tr}_{\mathcal{H}}\left(  \mathrm{tr}%
P-\langle\mathrm{tr}P\rangle_{\infty}\right)  . \label{Top-2}%
\end{equation}
In fact, we can easily confirm that the soliton solutions from (\ref{sLee-P})
to (\ref{newAS}) when substituted into (\ref{Top-2}) give the same value for
$Q$ calculated with (\ref{Top-1}).

\section{Noncommutative View of Electron System on the Plane}

In this section we consider a two dimensional system of electrons with spin in
the magnetic field $B$ perpendicular to the plane. Coulomb repulsive force
among the electrons is assumed. If we restrict the electron states to the
lowest Landau level, the system is reduced to that on the noncommutative plane
\cite{Pasquier:2000gz}\cite{Lee:2001fk}\cite{Magro:2003bs}.

Let us express the external magnetic field in terms of the vector potential
\begin{equation}
A_{x}=-\frac{By}{2},\ A_{y}=\frac{Bx}{2},
\end{equation}
where $x$ and $y$ are the coordinates on the plane. We can define the
independent oscillators in terms of the complex variables, $z=\frac{1}%
{\sqrt{2}}(x+iy)$ and $\bar{z}=\frac{1}{\sqrt{2}}(x-iy)$ as%
\begin{align}
a  &  =\theta\partial_{\bar{z}}+\frac{z}{2},\text{ }a^{\dagger}=-\theta
\partial_{z}+\frac{\bar{z}}{2},\nonumber\\
b  &  =\theta\partial_{z}+\frac{\bar{z}}{2},\text{ }b^{\dagger}=-\theta
\partial_{\bar{z}}+\frac{z}{2},
\end{align}
which satisfy
\begin{equation}
\lbrack a,a^{\dagger}]=[b,b^{\dagger}]=\theta,
\end{equation}
where $\theta=1/(eB)$ \cite{Pasquier:2000gz}\cite{Lee:2001fk}. The Hamiltonian
of the first quantized system is proportional to $Ba^{\dag}a$ . Thus, $a$ and
$a^{\dagger}$ induce the transitions among the Landau levels, while $b$ and
$b^{\dagger}$ induce the transitions within each Landau level.

As is well known, in the large $B$ limit, the system can be restricted to the
lowest Landau level (LLL), and the electrons in the LLL can be considered as a
system on the plane with the noncommutative coordinates $b$ and $b^{\dagger}$.
The Hilbert space is spanned by $\left\{  \left\vert n\right\rangle \right\}
$, for which
\begin{equation}
b^{\dagger}b\left\vert n\right\rangle =n\theta\left\vert n\right\rangle .
\end{equation}

The second quantized field that annihilates (creates) an electron with a spin
$\sigma$ at position $\vec{x}$ in the LLL can be constructed as
\begin{equation}
\Psi_{\sigma}\left(  \vec{x}\right)  =\sum_{n}\left\langle \vec{x}%
|n\right\rangle C_{n\sigma}\ ,\ \Psi_{\sigma}^{\dagger}=\left(  \Psi_{\sigma
}\right)  ^{\dagger}\ ,
\end{equation}
in terms of the fermionic operators $C_{n\sigma}$\textit{ }$\left(
C_{n\sigma}^{\dagger}\right)  $ which annihilates (creates) an electron in the
$n$-th orbital, where%
\begin{equation}
\left\langle \vec{x}|n\right\rangle =\left(  \frac{z}{\sqrt{\theta}}\right)
^{n}\frac{1}{\sqrt{2\pi n!}}\exp\left(  -\frac{\bar{z}z}{2\theta}\right)  \ .
\end{equation}
These satisfy the anticommutation relation%
\begin{equation}
\left\{  \Psi_{\sigma}\left(  \vec{x}\right)  ,\Psi_{\sigma^{\prime}}%
^{\dagger}\left(  \vec{x}\right)  \right\}  =\rho\delta_{\sigma\sigma^{\prime
}}\ ,
\end{equation}
where $\rho=\left\langle \vec{x}|\vec{x}\right\rangle =\left(  2\pi
\theta\right)  ^{-1}$. As we shall see, the system can be described in terms
of $P$-field that appeared in previous section which is expressed using the
electron field as%
\begin{equation}
P\mathbf{=}\frac{1}{\rho}\left(
\begin{array}
[c]{cc}%
\Psi_{\downarrow}^{\dagger}\Psi_{\downarrow} & \alpha\\
\alpha^{\dag} & \Psi_{\uparrow}^{\dagger}\Psi_{\uparrow}%
\end{array}
\right)  \mathbf{,}\text{ }P^{2}=P,
\end{equation}
where $\alpha$ can be an arbitrary function of $\Psi_{\sigma}$ and
$\Psi_{\sigma}^{\dagger}$ satisfying $P^{2}=P.$ Arbitrariness of $\alpha$ will
not play any role in the following discussions. Topological charge for the
case $\left\langle \mathrm{tr}P\right\rangle _{\infty}=1$ is%
\begin{equation}
Q=\frac{1}{2\pi\theta}\mathrm{Tr}_{\mathcal{H}}\left(  \mathrm{tr}%
P-\left\langle \mathrm{tr}P\right\rangle _{\infty}\right)  =\frac{1}%
{2\pi\theta}\mathrm{Tr}_{\mathcal{H}}\left(  \mathrm{tr}P-1\right)  ,
\end{equation}
which can be expressed in terms of the total number of magnetic fluxes
$N_{\phi}$\ and the number of electrons $N_{e}$ as%
\begin{equation}
Q={\frac{1}{2\pi\theta}}\mathrm{Tr}_{\mathcal{H}}\left(  \mathrm{tr}%
P-1\right)  =N_{e}-N_{\phi}.
\end{equation}
Here use has been made of%
\begin{equation}
\mathrm{tr}P=\frac{1}{\rho}\left(  \Psi_{\downarrow}^{\dagger}\Psi
_{\downarrow}+\Psi_{\uparrow}^{\dagger}\Psi_{\uparrow}\right)
\end{equation}
and%
\begin{equation}
N_{e}=\mathrm{Tr}_{\mathcal{H}}\left(  \Psi_{\downarrow}^{\dagger}%
\Psi_{\downarrow}+\Psi_{\uparrow}^{\dagger}\Psi_{\uparrow}\right)  .
\end{equation}
On the other hand, as $\left(  2\pi\theta\right)  ^{-1}$ is the number of
states per unit area of LLL, which is occupied by a unit flux, $N_{\phi}$ can
be considered to be the total number of states in the LLL. And the filling
factor is defined as the number of electrons per unit flux, $\nu=\frac{N_{e}%
}{N_{\phi}}$. Consequently, the topological number has a simple meaning of the
electron\ number added to (removed from) the state with filling factor $\nu=1$.

We consider the delta function like Coulomb repulsion potential between the
electrons, $V\left(  \left\vert z-z^{\prime}\right\vert \right)  =\rho
^{-1}\delta^{2}\left(  z-z^{\prime}\right)  $. Restricting the electron states
to the LLL, the Hamiltonian can be written as
\begin{align}
H  &  =\frac{1}{\rho}\int\left(  \Psi_{\uparrow}^{\dagger}\Psi_{\uparrow}%
-\Psi_{\downarrow}\Psi_{\downarrow}^{\dagger}\right)  ^{2}d^{2}x\nonumber\\
&  =\frac{2}{\rho}\int\left(  \Psi_{\uparrow}\Psi_{\downarrow}\right)
^{\dagger}\left(  \Psi_{\uparrow}\Psi_{\downarrow}\right)  d^{2}%
x+(-Q)\nonumber\\
&  =\rho\int\left(  \mathrm{{tr}}P-1\right)  ^{2}d^{2}x,
\end{align}
which is invariant under particle hole exchange \cite{Pasquier:2000gz}. Then
the energy is
\begin{equation}
E=\rho\mathrm{Tr}_{\mathcal{H}}\left(  \mathrm{{tr}}P-1\right)  ^{2}.
\label{H-electron}%
\end{equation}

We can compare our argument with that of ref.\cite{Pasquier:2000gz} which goes
as follows. The Hamiltonian for the above system written in terms of $\ast
$product is expanded in $\theta$ and this leads to the $O(3)\sigma$ model on
the commutative space. In this paper we shall discuss the problem of solitons
working directly within the operator formalism.

The energy inequality for $Q>0$ can be written as%
\begin{align}
E  &  =\rho\mathrm{Tr}_{\mathcal{H}}(\text{$\mathrm{tr}$}P-1)^{2}\nonumber\\
&  =\rho\mathrm{Tr}_{\mathcal{H}}\left\{  (\text{$\mathrm{tr}$}P-1)^{2}%
-(\text{$\mathrm{tr}$}P-1)\right\}  +\rho\mathrm{Tr}_{\mathcal{H}%
}(\text{$\mathrm{tr}$}P-1)\nonumber\\
&  =\rho\mathrm{Tr}_{\mathcal{H}}\left\{  (\text{$\mathrm{tr}$}P-1)^{2}%
-(\text{$\mathrm{tr}$}P-1)\right\}  +Q\nonumber\\
&  \geq Q,
\end{align}
where the first term on the third line is positive definite, as can be seen
from
\begin{equation}
\rho\mathrm{Tr}_{\mathcal{H}}\left\{  (\text{$\mathrm{tr}$}P-1)^{2}%
-(\text{$\mathrm{tr}$}P-1)\right\}  =\frac{2}{\rho}\mathrm{Tr}_{\mathcal{H}%
}\left\{  \left(  \Psi_{\downarrow}\Psi_{\uparrow}\right)  \left(
\Psi_{\downarrow}\Psi_{\uparrow}\right)  ^{\dagger}\right\}  \geq0.
\end{equation}
Consequently the BPS equation for $Q>0$ is%
\begin{equation}
(\text{$\mathrm{tr}$}P-1)^{2}=\text{$\mathrm{tr}$}P-1. \label{BPS-electron}%
\end{equation}
For $Q<0$ the energy inequality is%
\begin{align}
E  &  =\rho\mathrm{Tr}_{\mathcal{H}}(\text{$\mathrm{tr}$}P-1)^{2}\nonumber\\
&  =\rho\mathrm{Tr}_{\mathcal{H}}\left\{  (\mathrm{tr}P)^{2}-(\mathrm{tr}%
P)\right\}  -\rho\mathrm{Tr}_{\mathcal{H}}(\text{$\mathrm{tr}$}P-1)\nonumber\\
&  =\rho\mathrm{Tr}_{\mathcal{H}}\left\{  (\mathrm{tr}P)^{2}-(\mathrm{tr}%
P)\right\}  -Q\nonumber\\
&  \geq-Q
\end{align}
and we obtain the BPS equation,
\begin{equation}
(\mathrm{tr}P)^{2}=\mathrm{tr}P. \label{aBPS-electron}%
\end{equation}
The systems described by (\ref{L-nls}) and (\ref{H-electron}) are different,
thus the BPS eqs. (\ref{BPS-P}) (\ref{aBPS-P}) and (\ref{BPS-electron})
(\ref{aBPS-electron}) are different as they should be. As we shall see in the
following, however, the BPS eq. (\ref{BPS-P}) and (\ref{BPS-electron})
(anti-BPS eq. (\ref{aBPS-P}) and (\ref{aBPS-electron})) have common soliton
(anti-soliton) solutions.

An example of the BPS soliton for electron system with $Q=n>0$ is%
\begin{equation}
P=\left(
\begin{array}
[c]{cc}%
1 & 0\\
0 & \sum_{m=0}^{n-1}\left\vert m\right\rangle \left\langle m\right\vert
\end{array}
\right)  , \label{BPSE}%
\end{equation}
which has the energy $E=n$. An example of BPS anti-soliton with $Q=-n<0\ $is
\begin{equation}
P=\left(
\begin{array}
[c]{cc}%
1-\sum_{m=0}^{n-1}\left\vert m\right\rangle \left\langle m\right\vert  & 0\\
0 & 0
\end{array}
\right)  , \label{aBPSE}%
\end{equation}
which has the energy $E=n$. These soliton solutions are at the same time the
solutions of the nonlinear sigma model \cite{Otsu:2003fq}. On the other hand,
the following soliton solutions of the nonlinear sigma model
\cite{Lee:2000ey}
\begin{equation}
P=\left(
\begin{array}
[c]{cc}%
z^{n}\dfrac{1}{\bar{z}^{n}z^{n}+1}\bar{z}^{n} & z^{n}\dfrac{1}{\bar{z}%
^{n}z^{n}+1}\\
\dfrac{1}{\bar{z}^{n}z^{n}+1}\bar{z}^{n} & \dfrac{1}{\bar{z}^{n}z^{n}+1}%
\end{array}
\right)
\end{equation}
and%
\begin{equation}
P=\left(
\begin{array}
[c]{cc}%
\bar{z}^{n}\dfrac{1}{z^{n}\bar{z}^{n}+1}z^{n} & \bar{z}^{n}\dfrac{1}{z^{n}%
\bar{z}^{n}+1}\\
\dfrac{1}{z^{n}\bar{z}^{n}+1}z^{n} & \dfrac{1}{z^{n}\bar{z}^{n}+1}%
\end{array}
\right)
\end{equation}
do not satisfy the BPS equations (\ref{BPS-electron}) or (\ref{aBPS-electron}%
), and thus are not the solitons of the electron system, although these
configurations do have the finite energy.

Next, we comment on more general form of the soliton solutions in the electron
system. Non-BPS anti-soliton of ref.\cite{Furuta:2002nv}
\begin{equation}
P=\left(
\begin{array}
[c]{cc}%
1-\left\vert m\right\rangle \left\langle m\right\vert  & 0\\
0 & 0
\end{array}
\right)  \label{BPSinami}%
\end{equation}
is also a BPS soliton solution with $Q=-1$ and$\ E=1$. As far as
$\mathrm{tr}P$ is unchanged, arbitrary deformations of (\ref{BPSE}),
(\ref{aBPSE}) and (\ref{BPSinami}) are also the soliton solutions.

\section{Classification by New Topological Number}

In section 2, we have seen that the solitons of nonlinear $\sigma$ model on
the noncommutative plane are classified by the topological charge $Q$\ and the
new topological number $\left\langle \mathrm{tr}P\right\rangle _{\infty}$ .
Based upon this situation, in section 3, we have found that some solitons of
nonlinear $\sigma$ model with $\left\langle \mathrm{tr}P\right\rangle
_{\infty}=1$ can be considered as solitons in the model of electron system
with a filling factor $\nu\simeq1.$

What are the solitons with $\left\langle \mathrm{tr}P\right\rangle _{\infty
}=0,2$ ? As they have the topological charges
\begin{equation}
Q={\frac{1}{2\pi\theta}}\mathrm{Tr}_{\mathcal{H}}\left(  \mathrm{tr}P\right)
=N_{e}\text{ \ for }\left\langle \mathrm{tr}P\right\rangle _{\infty}=0
\end{equation}
and%
\begin{equation}
Q={\frac{1}{2\pi\theta}}\mathrm{Tr}_{\mathcal{H}}\left(  \mathrm{tr}%
P-2\right)  =N_{e}-2N_{\phi}\text{ \ for }\left\langle \mathrm{tr}%
P\right\rangle _{\infty}=2,
\end{equation}
respectively,\ the different topological numbers $\left\langle \mathrm{tr}%
P\right\rangle _{\infty}\mathrm{\ }$imply that these solitons belong to
different vacua with filling factors $\nu\gtrsim0,$ $\nu\lesssim2.$

Hamiltonians for the electron system corresponding to the solitons with
$\left\langle \mathrm{tr}P\right\rangle _{\infty}=0,2$ can be written as
\begin{align}
H_{0}  &  =\frac{1}{\rho}\int\left(  \Psi_{\uparrow}^{\dagger}\Psi_{\uparrow
}+\Psi_{\downarrow}^{\dagger}\Psi_{\downarrow}\right)  ^{2}d^{2}x\nonumber\\
&  =\rho\int\left(  \mathrm{{tr}}P\right)  ^{2}d^{2}x
\end{align}
and%
\begin{align}
H_{2}  &  =\frac{1}{\rho}\int\left(  \Psi_{\uparrow}\Psi_{\uparrow}^{\dagger
}+\Psi_{\downarrow}\Psi_{\downarrow}^{\dagger}\right)  ^{2}d^{2}x\nonumber\\
&  =\rho\int\left(  \mathrm{{tr}}P-2\right)  ^{2}d^{2}x,
\end{align}
respectively. We note, however, that the Coulomb repulsion could be negligibly
small for $\nu\approx0$, but for the sake of simplicity of our argument we
shall assume its existence for the whole region of $0\lesssim\nu\lesssim2$.
Then the energy for $\left\langle \mathrm{tr}P\right\rangle _{\infty}=0$ can
be expressed as
\begin{align}
E_{0}  &  =\rho\mathrm{Tr}_{\mathcal{H}}(\text{$\mathrm{tr}$}P)^{2}\nonumber\\
&  =\rho\mathrm{Tr}_{\mathcal{H}}\left\{  (\mathrm{tr}P)^{2}-(\mathrm{tr}%
P)\right\}  +\rho\mathrm{Tr}_{\mathcal{H}}(\text{$\mathrm{tr}$}P)\nonumber\\
&  =\rho\mathrm{Tr}_{\mathcal{H}}\left\{  (\mathrm{tr}P)^{2}-(\mathrm{tr}%
P)\right\}  +Q\nonumber\\
&  \geq Q,
\end{align}
which leads to the\ BPS equation%
\begin{equation}
(\mathrm{tr}P)^{2}=\mathrm{tr}P.
\end{equation}
In this case, we have solitons with $Q>0,\ $the concrete example of which with
$Q=n$ and $E=n$ is expressed by the projector%
\begin{equation}
P=\left(
\begin{array}
[c]{cc}%
\sum_{m=0}^{n-1}\left\vert m\right\rangle \left\langle m\right\vert  & 0\\
0 & 0
\end{array}
\right)  . \label{BPS0}%
\end{equation}
Similarly, the energy for $\left\langle \mathrm{tr}P\right\rangle _{\infty}=2$
is%
\begin{align}
E_{2}  &  =\rho\mathrm{Tr}_{\mathcal{H}}(\text{$\mathrm{tr}$}P-2)^{2}%
\nonumber\\
&  =\rho\mathrm{Tr}_{\mathcal{H}}\left\{  (\mathrm{tr}P-2)^{2}+(\mathrm{tr}%
P-2)\right\}  -\rho\mathrm{Tr}_{\mathcal{H}}(\text{$\mathrm{tr}$%
}P-2)\nonumber\\
&  =\rho\mathrm{Tr}_{\mathcal{H}}\left\{  (2-\mathrm{tr}P)^{2}-(2-\mathrm{tr}%
P)\right\}  -Q\nonumber\\
&  \geq-Q\ ,
\end{align}
and the BPS equation is%
\begin{equation}
(2-\mathrm{tr}P)^{2}=2-\mathrm{tr}P.
\end{equation}
Then there exist solitons with $Q<0,$ and the example of which with
$Q=-n\ $and $E=n$ is given by%
\begin{equation}
P=\left(
\begin{array}
[c]{cc}%
1-\sum_{m=0}^{n-1}\left\vert m\right\rangle \left\langle m\right\vert  & 0\\
0 & 1
\end{array}
\right)  . \label{aBPS2}%
\end{equation}
These soliton solutions (\ref{BPS0}) and (\ref{aBPS2}) are at the same time
the solutions of the nonlinear sigma model \cite{Otsu:2004fz}.

Thus, we have found that, as in the case of $\left\langle \mathrm{tr}%
P\right\rangle _{\infty}=1$, some solitons of the nonlinear sigma model with
$\left\langle \mathrm{tr}P\right\rangle _{\infty}=0,2$ can also be considered
to be the solitons of the electron system.

\section{Summary and Discussion}

In this paper, we have seen that some solitons of nonlinear $\sigma$ model on
the noncommutative plane can be considered as solitons in the model of two
dimensional electron system. This is rather surprising, because the BPS
equations for the electron system on one hand and the nonlinear $\sigma$ model
on the other are very different. These solitons are classified by two
topological numbers $Q$ and $\left\langle \mathrm{tr}P\right\rangle _{\infty
}=0,1,2$, as is described in section 2. Corresponding to these topological
numbers, in section 3 and 4, we have obtained the Hamiltonians, BPS equations
and soliton solutions of the electron system. The noncommutative solitons with
$\left\langle \mathrm{tr}P\right\rangle _{\infty}=0,1,2$ correspond to the
solitons with a filling factor $\nu\gtrsim0,\nu\simeq1$ and $\nu\lesssim2$, respectively.

It would be interesting to find the corresponding experimental evidence for
the soliton solutions of the electron system.

It is instructive to compare our noncommutative soliton solutions with those
of refs.\cite{Pasquier:2000gz}\cite{Lee:2001fk}, where they have arrived at
the $O(3)\sigma$ model on commutative space by expanding the Hamiltonian,
written in terms of the $\ast$product, in $\theta$. Thus their solitons are
those of commutative model. In present paper, on the other hand, we have
discussed the problem of solitons working directly within the operator
formalism. As we have seen, these solitons are different from the commutative
limits of our noncommutative solitons.

Finally, in connection with the recent work \cite{Domrin:2004pg} the stability
of our solutions have to be examined.

\providecommand{\href}[2]{#2}\begingroup \raggedright


\begin{thebibliography}{99}                                                                                               %


\bibitem {Harvey:2001yn}J.~A. Harvey, \textit{Komaba Lectures on
Noncommutative Solitons and D-Branes}, \texttt{hep-th/0102076}.

\bibitem {Nekrasov:1998ss}N.~Nekrasov and A.~Schwarz, \textit{Instantons on
Noncommutative }$R^{4}$\textit{ and }$(2,0)$\textit{ Superconformal Six
Dimensional Theory}, \emph{Commun. Math. Phys.} \textbf{198} (1998) 689--703,
[\texttt{hep-th/9802068}].

\bibitem {Gopakumar:2000zd}R.~Gopakumar, S.~Minwalla, and A.~Strominger,
\textit{Noncommutative Solitons}, \emph{JHEP} \textbf{05} (2000) 020,
[\texttt{hep-th/0003160}].

\bibitem {Lee:2000ey}B.-H. Lee, K.-M. Lee, and H.~S. Yang, \textit{The
CP}$^{n}$\textit{ Model on Noncommutative Plane}, \emph{Phys. Lett.}
\textbf{B498} (2001) 277--284, [\texttt{hep-th/0007140}].

\bibitem {Lechtenfeld:2001uq}O.~Lechtenfeld, A.~D. Popov, and B.~Spendig,
\textit{Noncommutative Solitons in Open n = 2 String Theory}, \emph{JHEP}
\textbf{06} (2001) 011, [\texttt{hep-th/0103196}].

\bibitem {Lechtenfeld:2001aw}O.~Lechtenfeld and A.~D. Popov,
\textit{Noncommutative Multi-Solitons in 2+1 Dimensions}, \emph{JHEP}
\textbf{11} (2001) 040, [\texttt{hep-th/0106213}].

\bibitem {Lechtenfeld:2001gf}O.~Lechtenfeld and A.~D. Popov,
\textit{Scattering of Noncommutative Solitons in 2+1 Dimensions}, \emph{Phys.
Lett.} \textbf{B523} (2001) 178--184, [\texttt{hep-th/0108118}].

\bibitem {Furuta:2002ty}K.~Furuta, T.~Inami, H.~Nakajima, and M.~Yamamoto,
\textit{Low-Energy Dynamics of Noncommutative CP}$^{1}$\textit{ Solitons in
2+1 Dimensions}, \emph{Phys. Lett.} \textbf{ B537} (2002) 165--172,
[\texttt{hep-th/0203125}].

\bibitem {Furuta:2002nv}K.~Furuta, T.~Inami, H.~Nakajima, and M.~Yamamoto,
\textit{Non-BPS Solutions of the Noncommutative CP}$^{1}$\textit{ Model in
(2+1)-Dimensions}, \emph{JHEP} \textbf{08} (2002) 009,
[\texttt{hep-th/0207166}].

\bibitem {Otsu:2003fq}H.~Otsu, T.~Sato, H.~Ikemori, and S.~Kitakado,
\textit{New BPS Solitons in 2+1 Dimensional Noncommutative CP}$^{1}$\textit{
Model}, \emph{JHEP} \textbf{07} (2003) 054, [\texttt{hep-th/0303090}].

\bibitem {Otsu:2004fz}H.~Otsu, T.~Sato, H.~Ikemori, and S.~Kitakado,
\textit{Lost Equivalence of Nonlinear Sigma and CP}$^{1}$\textit{ Models on
Noncommutative Space}, \emph{JHEP} \textbf{ 06} (2004) 006,
[{\texttt{hep-th/0404140}]. }

\bibitem {Pasquier:2000gz}V.~Pasquier, \textit{An Exact Correspondence between
Quantum Hall Skyrmions and Non Linear Sigma-Models}, \emph{Phys. Lett.}
\textbf{B513} (2001) 241--244, [\texttt{cond-mat/0012207}].

\bibitem {Lee:2001fk}B.-H. Lee, K.~Moon, and C.~Rim, \textit{Noncommutative
Field Theory Description of Quantum Hall Skyrmions}, \emph{Phys. Rev.}
\textbf{D64} (2001) 085014, [\texttt{hep-th/0105127}].

\bibitem {Magro:2003bs}G.~Magro, \textit{Noncommuting Coordinates in the
Landau Problem}, [{\texttt{quant-ph/0302001}]. }

\bibitem {Gopakumar:2001yw}R.~Gopakumar, M.~Headrick, and M.~Spradlin,
\textit{On Noncommutative Multi-Solitons}, \emph{Commun. Math. Phys.}
\textbf{233} (2003) 355--381, [\texttt{hep-th/0103256}].

\bibitem {Hadasz:2001cn}L.~Hadasz, U.~Lindstrom, M.~Rocek, and R.~von Unge,
\textit{Noncommutative Multisolitons: Moduli Spaces, Quantization, Finite
Theta Effects and Stability}, \emph{JHEP} \textbf{06} (2001) 040,
[\texttt{hep-th/0104017}].

\bibitem {Domrin:2004pg}A.~V. Domrin, O.~Lechtenfeld, and S.~Petersen,
\textit{Sigma-Model Solitons in the Noncommutative Plane: Construction and
Stability Analysis}, [\texttt{hep-th/0412001}].
\end{thebibliography}
\end{document}